
\input phyzzx.tex
\input tables.tex

\def\gam{\gamma}
\def\ptr{p_T}
\def\gev{~{\rm GeV}}

\def\fbi{~{\rm fb}^{-1}}

\def\smv{{\it Proceedings of the 1990 DPF Summer Study on
High Energy Physics: ``Research Directions for the Decade''},
editor E. Berger, Snowmass (1990)}

\def\prdj#1{{\it Phys. Rev.} {\bf D{#1}}}
\def\npbj#1{{\it Nucl. Phys.} {\bf B{#1}}}
\def\prlj#1{{\it Phys. Rev. Lett.} {\bf {#1}}}
\def\plbj#1{{\it Phys. Lett.} {\bf B{#1}}}

\def\mt{m_t}

\def\rta{\rightarrow}

\def\hsm{H}

\def\mhsm{m_{\hsm}}

\def\mw{m_W}

\def\anti{\overline}

\def\ifmath#1{\relax\ifmmode #1\else $#1$\fi}

\def\3quarter{{\textstyle{3 \over 4}}}

\def\ebtag{e_{b-tag}}
\def\emistag{e_{mis-tag}}

\input phyzzx
\Pubnum={$\caps UCD-93-18$\cr $\caps SMU-HEP-93/07$ \cr
$\caps SLAC-PUB-6265$\cr}
\date{June, 1993 (revised Sept. 1993)}

\titlepage
\vskip 0.75in
\baselineskip 0pt
\hsize=6.5in
\vsize=8.5in
\centerline{{\bf Using $\bf b$-Tagging to Detect $\bf t\anti t$~Higgs
Production
with Higgs$\bf\rta b\anti b$}}
\vskip .075in
\centerline{J. Dai$^a$, J.F. Gunion$^b$ and R. Vega$^c$}
\vskip .075in
\centerline{\it a) Dept. of Physics and Astronomy, Rutgers University,
Piscataway, NJ 08855}
\centerline{\it b) Davis Institute for High Energy Physics,
Dept. of Physics, U.C. Davis, Davis, CA 95616}
\centerline{\it c) Dept. of Physics, Southern Methodist University,
Dallas, TX 75275}
\centerline{\it and}
\centerline{\it Stanford Linear Accelerator Center, Stanford, CA 94305}

\vskip .075in
\centerline{\bf Abstract}
\vskip .075in
\centerline{\Tenpoint\baselineskip=12pt
\vbox{\hsize=12.4cm
\noindent We demonstrate that expected efficiencies and purities
for $b$-tagging at SSC/LHC detectors may allow detection of
the Standard Model Higgs in $t\anti t \hsm$ production, with $\hsm\rta b\anti
b$
decay, for $80\lsim\mhsm\lsim 130\gev$, provided $\mt\gsim 130\gev$.}}

\vskip .15in
\noindent{\bf 1. Introduction}
\vskip .075in

Understanding the Higgs sector is one
of the fundamental missions of future high energy colliders such
as the SSC and LHC. However, options for
detection of the Standard Model (SM) Higgs boson, $\hsm$, are limited
if $\mhsm$ lies in the low end of the intermediate mass range,
$80\lsim\mhsm\lsim 130\gev$, \ie\ below the region for which the
gold-plated $\hsm\rta Z^* Z\rta 4\ell$ is robustly viable
\REF\gkw{Originally proposed in
J.F. Gunion, G.L. Kane and J. Wudka, \npbj{299} (1988) 231.}
\refmark{\gkw}\ but above
the guaranteed mass reach of LEP-II. The most discussed detection modes
in this region rely entirely on the rare $\gam\gam$ decay channel of the $\hsm$
using either the inclusive $\hsm\rta\gam\gam$ mode \refmark{\gkw}\ or the
$l\gam\gam X$ final state arising from $W\hsm$
\REF\kks{R. Kleiss, Z. Kunszt, and J. Stirling, \plbj{253} (1991) 269.}
\REF\mangano{M. Mangano, SDC Collaboration Note SSC-SDC-90-00113.}
\REF\overview{J.F. Gunion, G.L. Kane \etal, {\it Overview and Recent
Progress in Higgs Boson Physics at the SSC}, \smv, p. 59.}
\refmark{\kks,\mangano,\overview}\
and $t\anti t \hsm$
\REF\gghttgunion{J.F. Gunion, \plbj{261} (1991) 510.}
\REF\gghttmarciano{W. Marciano and F. Paige, \prlj{66} (1991) 2433.}
\refmark{\gghttgunion,\gghttmarciano}\
associated production followed by $\hsm\rta\gam\gam$ decay.
Clearly, the establishment of viable
techniques for detection of the $\hsm$ in its main decay mode in this mass
region, $\hsm\rta b\anti b$, would be highly desirable.
In this letter, we demonstrate that if the top quark is not too light,
then expected $b$-tagging efficiencies and purities at the SSC/LHC
may enable one to obtain viable $\hsm$ signals
in the $b\anti b$ decay mode when the $\hsm$ is produced in association
with $t\anti t$.

A priori, many possible procedures can be envisioned for detection
of $t\anti t\hsm$ events. Of these, we have examined the four
which appear to be most worthy of investigation. In (1) we require that
both $t$'s decay to $\ell\nu b$ and search for the $\hsm$
as a peak in the $2j$ mass spectrum of
$\ell\ell 4j$ final states without $b$-tagging. In (2)-(4)
we require that three or four $b$'s be tagged
and look for a peak in the $2b$ mass spectrum.
If only two $b$'s are tagged, there is little gain against the enormous
background from $t\anti t $ related processes which automatically
yield two $b$'s in the final state.
By requiring that three or more $b$'s be tagged, such backgrounds can be
reduced to the same level as the irreducible background from $t\anti t
b\anti b$ production. In procedure (2) we require that both $t$'s decay
to $\ell\nu b$ and tag three $b$-jets.
In procedures (3) and (4), we only require that one (or both) of the $t$'s
decay
to $\ell\nu b$, and tag three or four $b$-jets, respectively.
In comparing (3) and (4) we shall see
that tagging four (or more) $b$'s may prove preferable.
The event rate is reduced, but the signals are
cleaner and statistically not that much worse.

Regarding procedure (1), in which no
$b$'s are tagged, it is absolutely necessary to specifically demand
four energetic jets in the final state in order to eliminate $t\anti t$
and $t\anti t g$ backgrounds.  The major backgrounds are then
$t\anti t gg$ and $t\anti t q\anti q$ ($q=u,d,s,c,b$).  An exact computation
of $t\anti t gg$ is not currently available. In our study,
we computed with exact matrix elements the $t\anti t\hsm$ signal,
and the $t\anti t Z$ and $t\anti t q\anti q$ backgrounds.
The resulting signal was of only marginal significance, even
though no $t\anti t gg$ background was included.

\REF\otherguys{T. Garavaglia, W. Kwong, and D.-D. Wu,
preprint SSCL-PP-189 (1993).}
This same mode has also been studied in the related work
of Ref.~\otherguys. There, an approximate computation of
the $t\anti t\hsm$ signal, and the $t\anti t Z$ and
$t\anti t gg$ backgrounds is performed with semi-optimistic results.
The $t\anti t q\anti q$ backgrounds were not computed.  In order to
compare to the work of Ref.~\otherguys, we repeated our calculations
with exactly their cuts. For $\mt=140\gev$, in the 10 GeV bin
around $\mhsm=100\gev$ we found a smaller signal rate
of $S=221$ events per SSC year, and background rate (excluding
$t\anti t gg$ events) of $B=2400$ events per SSC year, implying
$S/\sqrt{B}\sim 4$. (QCD ``K'' corrections factors of $K=1.6$
are included in these rates, but were not included in Ref.~\otherguys.)
The $t\anti t gg$ background is likely to at least double $B$, yielding
a very marginal significance for the $\hsm$. Thus, our conclusions for the
$2\ell 4j$ mode are considerably more pessimistic than those of
Ref.~\otherguys.

The only means for suppressing the large $t\anti t$ backgrounds is
to employ $b$ tagging. Indeed, only the irreducible backgrounds from
$t\anti t Z$ (with $Z\rta b\anti b$) and $t\anti t b\anti b$
processes would remain for sufficiently high efficiency and purity
of $b$-tagging.  In choosing canonical values for the latter
to employ in procedures (2)-(4) we have
been guided by the results obtained by the SDC collaboration.
\Ref\sdctdr{Solenoidal Detector Collaboration Technical Design Report,
E.L. Berger \etal, Report SDC-92-201, SSCL-SR-1215, 1992, p 4.15-4.16.}
These first studies achieved an efficiency for $b$-jet tagging via vertexing
alone that increases from about 25\% at $\ptr=30\gev$ to above 30\%
for $\ptr>40\gev$ for $b$-jets with $|\eta|<2-2.5$ (for $|\eta|\sim 2.5$
edge effects begin to reduce the efficiency).
Tagging a $b$-jet by a `high'-$\ptr$ lepton decay or neural network
was not included.
Under these same conditions, the probability of misidentifying a light
quark or gluon jet (that does not explicitly decay to $b\anti b$)
as a $b$-jet was found to be of order 1\%,
while that of mis-tagging a $c$ jet is about 5\%.

We have adopted the following as our `standard' procedure at the parton level.
To trigger on the events of interest, we require that at least one
of the $t$ quarks decay to an isolated lepton ($e$ or $\mu$).
At the parton level, the trigger lepton is required to have $\ptr\geq 20\gev$
and $|\eta|<2.5$ and to be isolated by
$\Delta R\geq 0.3$ from all other jets or leptons.
The $W$ from the other $t$ quark is allowed to decay
either leptonically or hadronically.
If both $W$'s decay leptonically, then we allow for either of the leptons
to provide the required isolated lepton trigger. Next,
any $b$-jet with $|\eta|<2$ and $\ptr>30\gev$
is assumed to have a probability of 30\% (independent of $\ptr$) to
be vertex tagged, provided there is no other vertex within $\Delta R_V=0.5$.
The probability to tag (\ie\ misidentify)
a light quark or gluon jet as a $b$-jet
under these same conditions is assumed to be 1\%.  Jets with $\ptr$
below $30\gev$ are assumed not to be tagged. Finally,
in order for a $b$-jet to be included in our invariant mass distributions
we require that it be separated from all other jets
(including other $b$'s) by at least $\Delta R_C=0.7$.

Our choices for $\Delta R_V$ and $\Delta R_C$ are important and
will be discussed more in a longer paper. In fact,
it may be that our choice of $\Delta R_V$ is too
conservative and that $b$-jets could be resolved for substantially smaller
$\Delta R_V$ separations.
\Ref\seiden{We thank A. Seiden and B. Hubbard of the SDC
collaboration for discussions on this issue.}
The most appropriate choice for $\Delta R_C$ is a particularly complex issue
that will require complete detector simulations to fully resolve.
If the energy of a tagged $b$-jet from the Higgs decay
cannot be accurately determined by calorimetry due to an overlapping
jet, then the mass of the $b\anti b$ pair will
not be determined with good accuracy.  Thus, we should not
insert into our $m(b\anti b)$ distribution those tagged
$b$'s (or $\anti b$'s) that are not sufficiently separated
from other jets (including other $b$'s).  This is experimentally
possible since the vertex of the $b$-jet will be visible and
a neighboring jet will be apparent via its energetic charged tracks that
do not track to the tagged $b$-jet vertex.
The normal standard of separation to avoid calorimetric
overlap is $\Delta R_C=0.7$.\refmark\sdctdr\
In this letter we consider only this value, although a variety of
considerations suggest that $\Delta R_C=0.5$ may also be viable.

Of course, it is important to explore sensitivity to our assumptions regarding
$b$-tagging efficiency and purity. For instance,
we shall demonstrate how much easier $\hsm$ detection becomes if
the $b$-tagging probability can be brought up to 40\%
while decreasing the misidentification
probability to 0.5\% over the stated kinematic range.
More detailed analysis techniques and
technical improvements in vertex tagger designs could
significantly increase the efficiency with which $b$-jets are tagged.
Meanwhile, the misidentification probability can probably also be reduced
by a closer examination of the tagged jets; no analysis
of the jets associated with a tagged vertex was performed in
obtaining the 1\% probability quoted in Ref.~\sdctdr.
For example, neural-net analyses of vertex tagged jets
could remove some of the mis-tagged light quark and gluon jets.

The final critical ingredient in our analysis is the mass resolution
that can be achieved for the combined mass of two $b$-jets.
For a purely hadronic jet the SDC TDR\refmark\sdctdr\
quotes energy resolutions (depending upon design and integration time)
that are typically no worse than $\delta E/E=0.5/\sqrt{E(GeV)}\oplus 0.03$
(leading to roughly an 8\% jet-pair invariant mass resolution
for $30\gev$ jets). For leptons the energy resolution is typically of order
$\delta E/E=0.2/\sqrt{E(GeV)}\oplus 0.01$. The first
step of our parton-level analysis is to explicitly
smear the energies of all leptons and jets using these resolutions. In
particular, our $m(b\anti b)$ distributions will automatically reflect the
jet energy resolution. Since $b$-jets have hard fragmentation functions
the above hadronic resolution may be somewhat pessimistic for purely
hadronic $b$ decays. However, semi-leptonic $b$ decays will have worse
resolution. For the results to be presented here,
we have used standard hadronic resolutions as a compromise.

Two distinct more detailed approaches to the semi-leptonic $b$ decays are
possible.  First, it might be experimentally possible to greatly limit the
presence of such decays by requiring that a $b$-vertex not have an associated
lepton with energy larger than some appropriate lower bound,
especially one with significant $p_T$ with respect
to the primary $b$-jet direction (as determined by the vertex location
and the primary collision point). Our results would then be unaltered
if the $b$-tagging efficiency we employ is reinterpreted
as including the probability for a purely hadronic primary $b$ decay.
Alternatively, semi-leptonic decays can be explicitly included
in the analysis.  We have done this, retaining the above
hadronic and leptonic energy resolutions
for the $c$ and $\ell$ in $b\rta c\ell\nu$ (taking $BR(b\rta c\ell\nu)=0.1$,
$\ell=e,\mu,\tau$) and using only the combined $c$ and $\ell$ momenta in
constructing the $b\anti b$ mass for decays of this type. For all $\mhsm$,
the height of the signal peak is reduced and
the `background' from the signal reaction itself, that would be
estimated using nearby bins outside $\mhsm\pm10\gev$, increases somewhat.
Keeping all other efficiencies constant, one finds roughly
a doubling in the times (quoted later) required to see a signal
of a given statistical significance.
However, if $b\rta c\ell\nu$ decays are retained, a significant improvement
in the $b$-tagging efficiency will probably be possible by explicitly
looking for the associated lepton.  Expectations\refmark\seiden\
are that at least 5\% would be added to the 30-40\% tagging efficiency.
In the case of 4 $b$-tagging, this would restore much of the
statistical significance obtained prior to including $b\rta c\ell\nu$ decays.
This type of analysis will be pursued at greater length elsewhere.%
\Ref\paperii{J. Dai, J.F. Gunion, and R. Vega, in preparation.}

Of course, we have employed the SM predictions for the $\hsm\rta b\anti b$
branching ratio. At $\mhsm=80$, 100, 120, $140\gev$ we find
$BR(\hsm\rta b\anti b)=0.857$ ,0.842, 0.748, and 0.437, respectively.
These results include QCD corrections to the $q\anti q$ decay modes.
Not surprisingly, we will find that by $\mhsm=140\gev$ the branching ratio has
fallen sufficiently that $\hsm$ detection in the $b\anti b$
mode becomes fairly difficult.

\smallskip
\noindent{\bf 2. Procedure}
\smallskip

We have employed exact matrix element calculations for the
signal reaction $t\anti t\hsm$
\Ref\kunszt{A purely numerical computation of this process appeared in
Z. Kunszt, \npbj{247}, 339 (1984).} (with $\hsm\rta b\anti b$),
the irreducible backgrounds $t\anti t b\anti b$ and $t\anti t Z$
(with $Z\rta b\anti b$), and the reducible backgrounds
$t\anti t c\anti c$ (with one or two $c$'s mis-tagged as a $b$)
and $t\anti t(g)$ (with one or two of the non-$b$ jets
mis-tagged as a $b$). In the case of the $t\anti t b\anti b$
matrix element, agreement with the results of
\REF\stange{V. Barger, A.L. Stange and R.J.N. Phillips,
\prdj{44}, 1987 (1991).}
Ref.~\stange, for the cuts and procedures specified in the latter
reference, is only within a factor of two to four.
We have checked our matrix element very carefully, and in particular
find exact agreement (for any given configuration of the momenta)
in the massless limit with the result obtained in
\REF\gk{J.F. Gunion and Z. Kunszt, \plbj{159}, 167 (1985).
Note that, as pointed out in Ref.~\stange, some entries in the
color matrix Table 2 were accidentally omitted in the
written version of the paper; they were, however,
included in the numerical results presented there.}
Ref.~\gk. For all reactions we have included correlations in
the three-body decays of the top quarks. This turns out to be
important for the $t\anti t (g)$ background in which one or more
of the mis-tagged jets comes from the decay of a $W$.  Uncorrelated
decays would lead to a roughly 25\% overestimate of this background.

All the production reactions we consider are dominated by $gg$ collisions.
We have employed distribution functions for the gluons evaluated at
a momentum transfer scale given by the subprocess energy.  It is
well-known that QCD corrections are substantial for $gg$ initiated processes.
For example, for the $gg\rta t\anti t$ process
the QCD correction `K' factor has been found to be of the order
of 1.6 for our choice of scale.
\Ref\qcdcor{P. Nason, S. Dawson, and R.K. Ellis,
\npbj{303}, 607 (1988).  See also, W. Beenakker \etal, \prdj{40},
54 (1989).}
Computations of the `K' factors for the other reactions we consider
are not yet available in the literature. We will assume that they
are of the same magnitude.  Our precise procedures follow.
Rates for the $t\anti t\hsm$, $t\anti t Z$, $t\anti t b\anti b$
and $t \anti t c\anti c$ processes have been
multiplied by a QCD correction factor of 1.6.
In the case of $t\anti t (g)$ we have incorporated
the `K' factor as follows. We have generated events without
an extra gluon ($t\anti t$ events) and have also generated events
with an extra gluon ($t\anti t g$ events) requiring that
the $\ptr$ of the extra gluon be $>30\gev$.  For this cutoff
one finds $\sigma(t\anti t g)\sim 0.6 \sigma(t\anti t)$.  Thus,
if the two event rates are added together without cuts
an effective `K' factor of 1.6 is generated.  Explicitly allowing
for an appropriate number of $t\anti t g$ events is important in
properly estimating the background from this source due to
mis-tagged non-$b$ jets. Our procedure should yield an upper limit
on the number of events with an extra gluon having $\ptr>30\gev$
and therefore potentially vertex-taggable.
The gluon distribution functions we have employed are those of HMRS.
\Ref\hmrs{P.N. Harriman, A.D. Martin, W.J. Stirling, and R.G. Roberts,
\prdj{42}, 798 (1990).}
We have also repeated the $\mt=140\gev$ calculations for the
EHLQ\Ref\ehlq{E. Eichten, I. Hinchliffe, K. Lane, C. Quigg, {\it Rev. Mod.
Physics} {\bf 56}, 579 (1984).}\ and updated MRS
\Ref\mrs{A.D. Martin, R.G. Roberts, and W.J. Stirling,
preprint RAL-92-078 (1992).}
distribution functions.
Differences are typically at the 10-15\% level, except for
the MRS ${\rm D}-'$ choice. Relative to MRS ${\rm D}0'$, for example,
the gluon distribution is smaller in the $x\gsim 0.006$ region
that dominates our calculations, and we find roughly a 20-25\% suppression
in all event rates compared to the $\mt=140\gev$ results we quote below.

In analyzing a given event, we make no attempt to identify whether
or not a tagged $b$-jet is associated with a $t$ quark decay or
(when $\mhsm$ is not near $\mw$)
if a mis-tagged light quark jet can be combined with another jet
to yield an invariant mass near $\mw$.
(Further improvements in the signal significances that we
obtain could be made if an efficient manner for doing either can be found.)
As a result, in our analysis we must consider all pairs of tagged $b$-jets
(or mis-tagged non-$b$-jets) as potentially coming from the $\hsm$ decay.
Consequently, even the signal process has a combinatoric background
arising from pairs in which one or both of the $b$'s come from
$t$ decays.  The event rate for the $\hsm$ signal is obtained
by subtracting this combinatoric background from the $\hsm$ peak
in the $m(b\anti b)$ mass distribution. The background rate we employ
includes this signal combinatoric background as well as the full
backgrounds (including all combinatorics) for the true irreducible
and reducible backgrounds.  The number of
$b\anti b$ pairs that are included in the combinatoric backgrounds
is determined on an event by event basis according to the appropriate
probabilities for tagging or mis-tagging $b$ or non-$b$ jets,
respectively.


We shall present detailed
results for two $b$-tagging scenarios specified by giving:
i) the efficiency for tagging a $b$-jet, $\ebtag$; and ii)
the probability of mis-tagging a light quark or gluon jet, $\emistag$.
In case a), b) we take $(e_{b_tag},\emistag)=(30\%,1\%)$, $(40\%,0.5\%)$,
respectively. In both cases we have taken the probability of mis-tagging a
$c$-jet as being 5\%.  In fact, for mis-tagging probabilities of this general
level the $t\anti t c\anti c$ background is not significant.
(This is because the $t\anti t c \anti c$ and $t\anti t b\anti b$ cross
sections are not very different once high $\ptr$ is required for the
$c$'s or $b$'s, respectively.) Case a) we regard as a lower bound, and
case b) we think is quite achievable. For further comparison,
we shall also give SSC results at
$\mt=140\gev$ obtained using the $\ptr$-dependent
$\ebtag$ values of Ref.~\sdctdr\ for $\ptr>20\gev$, for
$\emistag=0.5\%$, $1\%$, and $1.5\%$.

\smallskip
\noindent{\bf 3. Results and Discussion}
\smallskip

\FIG\massplotthreeb{}
\topinsert
\vbox{\phantom{0}\vskip 4.5in
\phantom{0}
\vskip .5in
\hskip +30pt
\special{ insert scr:hbb_sm_fig2.ps}
\vskip -1.4in }
\centerline{\vbox{\hsize=12.4cm
\Tenpoint
\baselineskip=12pt
\noindent
Figure~\massplotthreeb: Events per 1 GeV bin in $m(b\anti b)$ for
signal plus background (solid) compared to background alone (dashes).
Results for the SSC with $L=10\fbi$,
and $b$-tagging efficiency and purity given by
$(\ebtag,\emistag)=(30\%,1\%)$ (lower histograms)
and $(40\%,0.5\%)$ (upper histograms) are shown for 3 $b$-tagging. Note that
the lower histograms for the former case have been shifted downwards
by 100 events per bin in order to clarify the display.
The Higgs signals at $m(b\anti b)=80$, $100$, $120$, and $140\gev$
are displayed after removing the combinatoric background
from the $t\anti t \hsm$ reaction itself.
See discussion in the text for details.
}}
\endinsert

\FIG\massplotfourb{}
\topinsert
\vbox{\phantom{0}\vskip 4.5in
\phantom{0}
\vskip .5in
\hskip +30pt
\special{ insert scr:hbb_sm_fig1.ps}
\vskip -1.4in }
\centerline{\vbox{\hsize=12.4cm
\Tenpoint
\baselineskip=12pt
\noindent
Figure~\massplotfourb: Events per 5 GeV bin in $m(b\anti b)$ for
signal plus background (solid) compared to background alone (dashes).
Results for the SSC with $L=10\fbi$,
and $b$-tagging efficiency and purity given by
$(\ebtag,\emistag)=(30\%,1\%)$ (lower histograms)
and $(40\%,0.5\%)$ (upper histograms) are shown for 4 $b$-tagging.
The Higgs signals at $m(b\anti b)=80$, $100$, $120$, and $140\gev$
are displayed after removing the combinatoric background
from the $t\anti t \hsm$ reaction itself.
See discussion in the text for details.
}}
\endinsert

In this letter, we present results for procedures (3) and (4)
in which only one lepton is required for triggering and
either three (or more) or four (or more), respectively,
jets must be tagged as $b$-jets. Procedure (2), in which two leptons
are triggered on and three jets are required to be tagged as $b$'s
yields smaller statistical significance for the Higgs signals
than either procedure (3) or (4).
Only for a Higgs mass near $\mw$ could it be of significant utility
by virtue of the absence of the peak near $\mw$ that can arise
if both of the jets from a hadronic $W$ decay are mis-identified
as $b$-jets.

In order to give a rough idea of the levels of the Higgs signals
and the various backgrounds we first present plots of event
rates as a function of $m(b\anti b)$ in Figs.~\massplotthreeb\
and \massplotfourb, for procedures (3) and (4), respectively, at $\mt=140\gev$.
In each figure, the two $b$-tagging $(\ebtag,\emistag)$
cases of (30\%,1\%) and (40\%,0.5\%) are compared.
All jet energies have been smeared using the resolution quoted earlier.
Results presented are for the SSC with integrated luminosity of $L=10\fbi$.
Since the combinatoric background from the $t\anti t \hsm$ reaction
itself is $\mhsm$ dependent, we have adopted the following procedure
for displaying all the Higgs mass peaks on one plot. For each $\mhsm$
case, we have taken only the bins that lie within $\pm 10\gev$ of
the Higgs peak in the $m(b\anti b)$ distribution.
{}From the event numbers in each of these central bins we have
subtracted an approximate combinatoric background determined by
averaging the distribution values for a representative set of
bins immediately below and immediately above the central bins.
The remainder in each of the central bins is then added
to the event number distributions coming from the sum of all other processes.
Thus, the upper (solid) histogram corresponds to the sum of event rates given
by
$$(N(t\anti t\hsm)-N(t\anti t\hsm)_{\rm comb.})+N(t\anti t Z)
+N(t\anti t) +N(t\anti t g)+N(t\anti t b\anti b)\,.\eqn\sum$$
To obtain the actual background
level in the vicinity of each peak one must add in the signal combinatoric
background appropriate to that value of $\mhsm$.  All other
combinatoric effects are included in the backgrounds incorporated
in the figures. For the scenarios illustrated, the $t\anti t\hsm$
combinatoric background, $N(t\anti t\hsm)_{\rm comb.}$,
is roughly 1/4 of the average value of
$N(t\anti t\hsm)-N(t\anti t\hsm)_{\rm comb.}$ in the bins
within $\pm 5\gev$ of the Higgs peak.
Also histogrammed is the event rate for the $t\anti t Z+
t\anti t+t\anti t g+t\anti t b\anti b$ background.
Additional graphs of signal and individual backgrounds
as a function of $m(b\anti b)$ will appear elsewhere.%
\refmark\paperii\

 \TABLE\ssclum{}
 \topinsert
 \titlestyle{\twelvepoint
 Table \ssclum: Number of $10\fbi$ years (signal event rate)
 at the SSC required for a  $5\sigma$ confidence level signal in four cases:
 I), II) --- 3 $b$ tagging with $(\ebtag,\emistag)=(30\%,1\%)$,
$(40\%,0.5\%)$; and
 III), IV) --- 4 $b$ tagging with $(\ebtag,\emistag)=(30\%,1\%)$,
$(40\%,0.5\%)$.}
 \bigskip

 \thicksize=0pt
 \hrule \vskip .04in \hrule
 \begintable
 Case | $\mt$\\$\mhsm$ | 80 & 100 & 120 & 140 \cr
 \ | 110 |   2.1( 331) &   3.5( 411) &   6.0( 458) &  29.1( 845) \nr
 I | 140 |   1.4( 324) &   3.0( 512) &   4.5( 486) &  22.4( 954) \nr
 \ | 180 |   0.3(  96) &   0.8( 175) &   2.9( 353) &  13.6( 747) \cr
 \ | 110 |   0.5( 191) &   1.0( 249) &   1.6( 275) &   7.6( 522) \nr
 II | 140 |   0.3( 170) &   0.6( 243) &   1.1( 266) &   5.3( 508) \nr
 \ | 180 |   0.1(  48) &   0.2(  97) &   0.6( 171) &   3.1( 375) \cr
 \ | 110 |  10.8( 135) &  15.9( 144) &  33.0( 186) & 170.0( 386) \nr
 III | 140 |   3.5(  95) &   5.0(  97) &  10.4( 122) &  55.0( 247) \nr
 \ | 180 |   1.0(  39) &   1.7(  45) &   4.2(  68) &  25.5( 160) \cr
 \ | 110 |   2.4(  94) &   4.1( 116) &   8.6( 153) &  46.0( 330) \nr
 IV | 140 |   0.6(  54) &   1.1(  66) &   2.6(  96) &  13.3( 189) \nr
 \ | 180 |   0.2(  23) &   0.4(  33) &   0.9(  46) &   5.5( 110) \endtable
 \hrule \vskip .04in \hrule
 \endinsert

 \TABLE\lhclum{}
 \topinsert
 \titlestyle{\twelvepoint
 Table \lhclum: Number of $100\fbi$ years (signal event rate)
 at the LHC required for a  $5\sigma$ confidence level signal in four cases:
 I), II) --- 3 $b$ tagging with $(\ebtag,\emistag)=(30\%,1\%)$,
$(40\%,0.5\%)$; and
 III), IV) --- 4 $b$ tagging with $(\ebtag,\emistag)=(30\%,1\%)$,
$(40\%,0.5\%)$.}
 \bigskip

 \thicksize=0pt
 \hrule \vskip .04in \hrule
 \begintable
 Case | $\mt$\\$\mhsm$ | 80 & 100 & 120 & 140 \cr
 \ | 110 |   1.5( 407) &   2.4( 488) &   6.0( 654) &  27.2(1146) \nr
 I | 140 |   0.8( 365) &   2.0( 600) &   3.5( 621) &  21.7(1338) \nr
 \ | 180 |   0.2( 102) &   0.6( 210) &   1.8( 373) &  12.5( 963) \cr
 \ | 110 |   0.4( 229) &   0.6( 283) &   1.5( 368) &   6.6( 662) \nr
 II | 140 |   0.2( 179) &   0.4( 291) &   0.8( 323) &   4.7( 690) \nr
 \ | 180 |   0.1(  62) &   0.1( 109) &   0.4( 195) &   2.3( 422) \cr
 \ | 110 |   6.4( 145) &   9.9( 151) &  22.4( 204) & 111.9( 395) \nr
 III | 140 |   2.0(  99) &   3.3( 112) &   7.5( 139) &  44.3( 303) \nr
 \ | 180 |   0.7(  43) &   1.2(  49) &   2.9(  74) &  18.8( 176) \cr
 \ | 110 |   1.3(  92) &   2.4( 116) &   5.6( 160) &  28.7( 321) \nr
 IV | 140 |   0.3(  51) &   0.7(  71) &   1.8( 103) &  10.5( 227) \nr
 \ | 180 |   0.1(  25) &   0.3(  35) &   0.6(  48) &   3.9( 114) \endtable
 \hrule \vskip .04in \hrule
 \endinsert

Several features of Figs.~\massplotthreeb\ and \massplotfourb\
are worth noting.  First, it is evident that the $m(b\anti b)$ peak from
the $t\anti tZ$ background is quite small.
(This means that the $Z$ peak from $t\anti tZ$ will
be quite difficult to detect in the $m(b\anti b)$ spectrum.)
Second, it is clear that the 4 $b$-tagging graph with
$(\ebtag,\emistag)=(40\%,0.5\%)$ has the cleanest signal peaks.
This is because the $t\anti t (g)$ backgrounds have been all but eliminated,
leaving primarily the irreducible $t\anti t b\anti b$ background.
If good efficiency and, especially, purity cannot be achieved,
for 4 $b$-tags the $t\anti t (g)$ backgrounds are comparable to
the $t\anti t b\anti b$ background, and for 3 $b$-tags they are dominant.
Similar plots for $\mt=180$ exhibit quite dramatic Higgs peaks for
all but the $\mhsm=140\gev$ case.

In order to quantify the observability of the Higgs signals,
such as those illustrated in Figs.~\massplotthreeb\ and \massplotfourb,
we have computed the number of SSC or LHC years required for a $5\sigma$
significance of the signal, defined by
$S\equiv N(t\anti t\hsm)-N(t\anti t\hsm)_{\rm comb.}$, summed over
bins within $\pm 5\gev$ of a given Higgs mass peak.
The background at each value of $\mhsm$
is computed as $B\equiv N(t\anti t\hsm)_{\rm comb.}+N(t\anti t Z)
+N(t\anti t) +N(t\anti t g)+N(t\anti t b\anti b)$ summed over
these same central bins. Results are given in Tables~\ssclum\ and \lhclum,
respectively.  In each table results for $b$-tagging scenarios a) and b)
are displayed for both 3 and 4 $b$-tagging.
Also given (in parentheses) is the associated number of signal events
($S$).  The associated number of background events ($B$)
can be obtained from the relation $B=S^2/25$. So that trends
can be clearly illustrated,
we have not made an arbitrary cutoff in the number of years
allowed for our entries.  In addition, it should be kept in mind
that (at least at the SSC) the ultimate luminosity that can be achieved
and managed by the detectors might be much larger than currently assumed.

{}From these two tables it is immediately apparent that for the two larger
$t$ masses, especially for $\mt=180\gev$,
detection of the $\hsm$ using $b$-tagging should be possible
within a few SSC or LHC years for the more optimistic scenarios,
so long as $\mhsm\lsim 130\gev$.  By $\mhsm\gsim 140\gev$
the $\hsm\rta b\anti b$ branching ratio has dropped to too small
a value due to the onset of the $WW^*$ decay modes.
The $\mt=110\gev$ entries
illustrate the fact that $\hsm$ detection using $b$-tagging
will be difficult for lighter top quark masses. The $t\anti t (g)$
background is much larger and the signal rates somewhat smaller
than for the larger $\mt$ values considered.
We estimate that $\mt=130\gev$ is the boundary below which detection
difficulty increases dramatically. On a statistical basis,
3 $b$-tagging is superior to 4 $b$-tagging.  But this is only true
if the background shapes can be reliably Monte Carlo'd.  In the
absence of a reliable prediction for the background shape, 4 $b$-tagging
could prove the preferable procedure. Of course, the 3 $b$-tagging
and 4 $b$-tagging procedures are complementary and the data should be
analyzed in both ways.

For canonical luminosities the LHC is somewhat superior to the SSC
for a given scenario.  However,
the efficiency of $b$-tagging at the LHC, given the many overlapping
events expected, may not be as great as assumed here. Overlapping
events will also reduce the probability that the $b$'s from the Higgs
decay in a signal event will be isolated by $\Delta R_C$ from
other jets.

The rapid worsening of $\hsm$ detectability in the $t\anti t b\anti b$
channel as we move from optimistic to pessimistic $(\ebtag,\emistag)$ scenarios
illustrates the importance of having as good a
vertex tagging capability as possible at the SSC and LHC.
The minimum value of $\Delta R_C$ for which good separation of the calorimetric
energy of a $b$-jet from that of a neighboring jet can be achieved
is also critical.  For instance, for $\mt=140\gev$ at the SSC
the signal rate at $\mhsm=100\gev$
increases by about 40\% when $\Delta R_C$ is decreased from 0.7 to 0.5.
(The $t\anti t b\anti b$ background increases by a similar amount,
but the $t\anti t (g)$ background increases by only about 25\%.)
Clearly, performing refined detector simulation
studies for this mode to assess which best represents
the true situation is of great importance.

To further define the $b$-tagging efficiency and purity that will be required
for observable signals, we also give results obtained by
employing the $\ptr$-dependent $\ebtag$
values of Ref.~\sdctdr\ down to $\ptr>20\gev$. These $\ebtag$ values
are somewhat lower than $30\%$ for $\ptr\lsim 60\gev$, but reach $\sim40\%$
for $\ptr\gsim120\gev$.
The number of SSC years required for a $5\sigma$ signal at $\mt=140\gev$
for $\emistag=(0.5,1,1.5)\%$, and for $\mhsm=80$, $100$, $120$, and
$140\gev$, respectively, is:
\#SSC years $=(0.8,1.3,1.9)$, $(1.8,3.1,4.5)$, $(3.3,5.4,7.5)$,
and $(12.1,20.0,29.8)$ for $3b$-tagging, and $=(1.8,3.0,5.1)$, $(2.8,4.0,6.1)$,
$(6.5,8.2,11.0)$, and $(39.3,48.3,63.4)$ for $4b$-tagging.
Comparing the $\emistag=1\%$ numbers to those in Table~\ssclum\
(rows I and III) for the uniform $\ebtag=0.3$ value, we see little
change for $3b$-tagging and some improvement for $4b$-tagging.
The larger $\ptr$ range and higher $\ebtag$
values at large-$\ptr$ at the very least compensate for the
smaller $\ebtag$ values at moderate $\ptr$.
Comparing results for the three $\emistag$ cases
shows clearly how critical the achievable $b$-tagging purity is.
An average $\emistag$ value significantly above 1\% would make observation
quite difficult for $\mhsm\gsim 100\gev$.

As a final remark, we note that although the lowest $\mhsm$ value studied
is $80\gev$, our figures and tables make it clear that the
$\hsm\rta b\anti b$ detection
modes studied here are, if anything, more viable at still somewhat lower
$\mhsm$ values.  This is in sharp contrast to the $\hsm\rta\gam\gam$ discovery
modes which rapidly deteriorate (due to decreasing $BR(\hsm\rta\gam\gam)$)
for $\mhsm< 80\gev$.

\smallskip
\noindent{\bf 4. Conclusion}
\smallskip

We have demonstrated that at the SSC and LHC expected vertex
tagging capabilities for a typical detector should be sufficient to allow
$\hsm$ detection in the $t\anti t \hsm\rta \ell b\anti b b\anti b X$ final
state
for a range of larger $\mt$ values and moderate $\mhsm$.
The precise region of viability depends critically upon the
efficiency and purity of $b$-tagging, but, allowing
for several years of running, should extend
at least from $\mhsm=80$ (and below)
to $\mhsm\sim 110\gev$ for $\mt\geq 140\gev$.
For $\mt\sim 180\gev$ and good efficiency and purity, $\mhsm$ values
up to $\sim 130\gev$ can be probed in just one SSC year.

\smallskip\noindent{\bf 5. Acknowledgements}
\smallskip
This work has been supported in part by Department of Energy
grants \#DE-FG03-91ER40674 and \#DE-AC03-76SF00515,
and by Texas National Research Laboratory grants \#RGFY93-330
and \#RCFY93-229.
JFG is grateful to A. Seiden, B. Hubbard and S. Willenbrock
for helpful conversations.
JFG also benefitted from early discussions with T. Weiler.
JD would like to thank M. Peskin and T. Han for discussions, and
D. Wu for explaining the work of Ref.~\otherguys.
\smallskip
\refout
\end